\documentclass{aa}  

\usepackage{graphicx}
\usepackage{txfonts}
\usepackage{hyperref}
\hypersetup{
  colorlinks   = true, 
  urlcolor     = blue, 
  linkcolor    = blue, 
  citecolor   = blue 
}
\usepackage[flushleft]{threeparttable}

\begin{document}

   \title{Studying the X-ray absorption characteristics of Centaurus X-3 using nearly 14 years of \textrm{MAXI}/GSC data}


   \author{ Ajith Balu\inst{1},
            Kinjal Roy\inst{1},
            Hemanth Manikantan\inst{1},
            Abhisek Tamang\inst{1,2}
            and Biswajit Paul\inst{1}
          }

    \institute{ Raman Research Institute, C. V. Raman Avenue, Sadashivanagar, Bangalore, Karnataka - 560080, India\\
                \email{ajithb@rrimail.rri.res.in}
                \and
                Indian Institute of Science, C. V. Raman Avenue, Bangalore, Karnataka - 560012, India\\
             }

   \date{Received July xx, 2024; accepted July xx, 2024}

 
  \abstract
   {Centaurus X-3 is a persistent high-mass X-ray binary with the long-term light curve from the source exhibiting orbit-to-orbit intensity variations with no apparent superorbital periodicity.}
   {We used $\sim$13.5 years of \textrm{MAXI}/GSC data to study the long-term behaviour of X-ray absorption caused by the stellar wind from the companion star and any absorbing structures present in the binary.}
   {We used orbital-phase-resolved spectroscopy to study the variation in the photoelectric absorption along the line of sight of the source for both the intensity-averaged data and intensity-resolved data after dividing all the data binned with orbital period into three intensity levels.}
   {We find an asymmetric variation in the photoelectric absorption along the line of sight across an orbit of the source. The orbital-phase-resolved spectra show a clear increase in photoelectric absorption after $\phi_\text{orb}\sim$ 0.5, which deviates from a spherically symmetric stellar wind model. The flux of Cen X-3 shows significant variation between consecutive orbits. An intensity-resolved spectral analysis of the source was performed, followed by an intensity-resolved and orbital-phase-resolved spectral analysis, which showed that at the medium and high intensity levels, the orbital-phase-resolved photoelectric absorption is slightly asymmetric with respect to mid-phase ($\phi_\text{orb}=$ 0.5). The asymmetry is very pronounced at the lowest intensity level and cannot be explained by a spherically symmetric wind from the companion star.}
   {The differences in the orbital phase-dependence of absorption for different intensity levels suggest that the presence of an accretion wake, photoionization wake, or tidal stream is more prominent at a lower intensity level for Centaurus X-3 than at a higher intensity level.}


    \titlerunning{Long-term study of Cen X-3}
    \authorrunning{Balu, A. et al.}
    \maketitle

\section{Introduction}
    High-mass X-ray binaries (HMXBs) are binary systems with a massive main sequence companion star with a mass $\geq10\,M_{\odot}$ and a compact object, which can be a neutron star (NS), white dwarf, or black hole. The optical companions in HMXBs are massive OB-type stars, which are known to have strong stellar winds driven by line absorption \citep{mass_loss_from_hot_star,CAK_75_wind_model}. Some of the mass lost in this stellar wind is accreted onto the compact object after being captured by its gravity, thus giving rise to X-ray emission.
    
    The main sequence stars in HMXB systems can have strong stellar winds with a mass-loss rate as high as ${\sim10^{-7}-10^{-6}\,M_{\odot}\,\mathrm{yr}^{-1}}$ \citep{blondin_wind_sim_1990}. The matter lost can cause photoelectric absorption of X-ray photons from the NS along our line of sight. Stellar wind models are used to probe the stellar wind from an OB-type star in a binary system with an accreting NS using pointed source observations \citep{wind_model_cenx3_photoionization_wojdowski, wind_mod_ex_pulse_phase_res_anal_cenx3_2_orb} or long-term observations with an all-sky monitor \citep{wind_mod_ex_footprints_velax1_paper, gx301m2_maxi_nazma_paul_2014, wind_mod_ex_4U_1538_52_with_MAXI, gx301m2_maxi_hemanth_paul_2023}.
    
    Centaurus X-3 (Cen X-3) is an HMXB pulsar system with an O6-8 III giant star (V779 Cen) \citep{Krzeminski_1974_otype, hutchings_1979_otype} as its optical companion. The companion star has a mass of $20.2^{+1.8}_{-1.5}\,M_{\odot}$, while the NS has a mass of $1.34^{+0.16}_{-0.14}\,M_{\odot}$ \citep{van_der_meer_2007}. The NS has a spin period of ${\sim 4.82\text{ s}}$ and the orbital period of the system is ${\sim 2.1\text{ days}}$ \citep{Nagase_1989_orbit_period_table}.
    
    In this paper, we present a long-term study of Cen X-3 using the Monitor of All-sky X-ray Image (\textrm{MAXI}) / Gas Slit Camera (GSC) all-sky monitor \citep{maxi_main_paper}, which has been in operation for over a decade. Such an instrument is ideal for an orbital-phase-resolved spectral analysis \citep{cenx3_six_year_paper} of the source. Also, the source exhibits different intensity levels in the \textrm{MAXI}/GSC data. Therefore, we investigated the intensity level variation in Cen X-3 \citep{harsha_flux_variation} and the orbital phase dependence of the source spectrum at each intensity level. After this, we fitted the photoelectric absorption observed with a symmetric stellar wind model and inspected the deviation of the measurements from an isotropic wind model at different intensity levels.
    
    \begin{figure}
    \centering
    \includegraphics[width=0.85\linewidth]{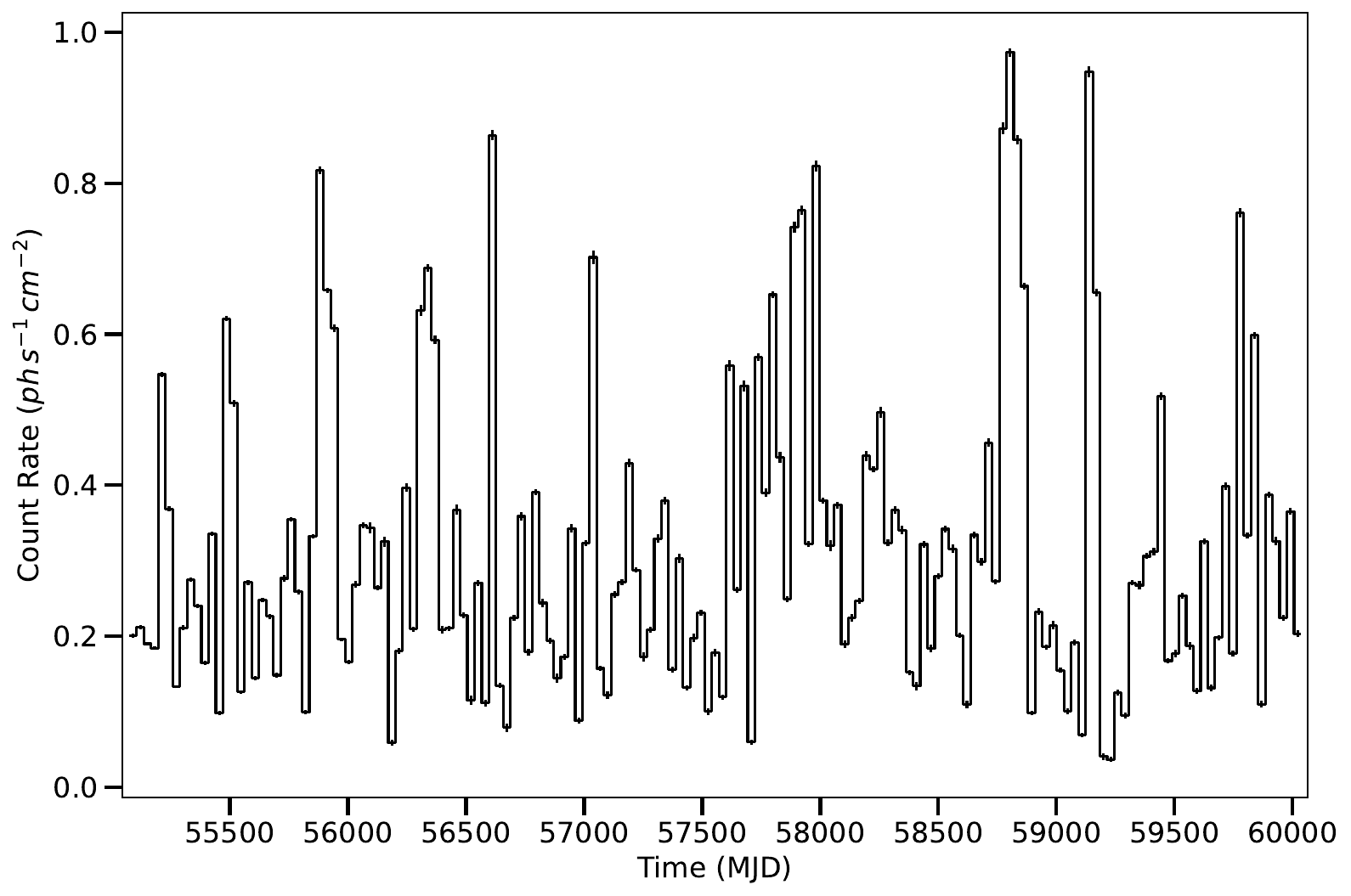}
    \caption{Long-term \textrm{MAXI}/GSC 2$-$20 $\mathrm{keV}$ light curve of Cen X-3 with a time binning of 30 days.}
    \label{fig:long_term_curve}
    \end{figure}
    
    Previous observations of several HMXBs \citep{para_accretion_wake_jackson, Kaper_photoionization_wake_1994} have shown the presence of additional structures around the compact object. These structures could be accretion wakes, photoionization wakes, or tidal streams \citep{blondin_wind_sim_1990, Blondin_1991, wind_loss_hydrodynamical_sim_2012}. Such structures lead to asymmetry in the orbital phase dependence of the absorption column density. In another study, long-term intensity variations were proposed to be caused by the accretion disc obscuring the compact object \citep{harsha_flux_variation}. For this work, we conducted orbital-phase-resolved spectral analysis for multiple intensity levels to thoroughly analyse differences in absorption as a function of the orbital phase for each intensity level.
    
    In Section \ref{timing_analysis}, we present the timing analysis we carried out, for which we divided the orbit into different regions after calculating the variation in hardness ratio ($HR$) from the soft and hard energy bands. In Section \ref{spectral_analysis}, we provide our orbital-phase-averaged and orbital-phase-resolved spectral analysis performed on both the intensity-averaged and intensity-resolved data. We then fitted a symmetric wind model for both the intensity-averaged and intensity-resolved data (Sect. \ref{sec:wind_modelling}) to investigate the variation in absorption for different intensity levels. A toy wake model was then used to understand the orbital phase dependence of absorption found for the lowest intensity level in Cen X-3.

\section{Observation and data}

    \begin{table}
        \caption{Orbital ephemeris of Cen X-3 from \citet{moritz_klawin_orbital_ephemeris_2023} and the orbital period calculated from $\sim$13.5 years of \textrm{MAXI}/GSC data.}
        \label{table:orbital_ephemeris}
        \renewcommand{\arraystretch}{1.2}
        \centering
        \begin{threeparttable}
        \setlength\tabcolsep{8.0pt}
        \begin{tabular}{ll}
            \hline
            Parameter                                                   & Value \\ \hline
            $E_{0}$ (MJD)                                & 40958.350335(26)  \\
            $P_\mathrm{orb}$ (days)                                       & 2.087139842(18)   \\
            $\Dot{P}_\mathrm{orb}$ (days days$^{-1}$)                     & -1.03788(27)$\,\times\,10^{-8}$   \\
            $P_\mathrm{orb,\,\textrm{MAXI}/GSC data}$ (days)                                       & 2.08697(2)   \\ \hline
        \end{tabular}
        \end{threeparttable}
    \end{table}
    
    Data from \textrm{MAXI} \citep{maxi_main_paper}, which has observed the whole sky every day for a period of approximately 14 years, were used in this study. \textrm{MAXI} is installed on the Japanese Experimental Module Exposed Facility (Kibo-EF) on board the International Space Station (ISS). As the ISS orbits the Earth, completing an orbit every ${\sim 90 \text{ minutes}}$, \textrm{MAXI} surveys the sky. Such observations allow for the study of the long-term variability of X-ray sources, which can be observed due to accretion disk instability, varying mass accretion rate, absorption, etc. The light curves for this study are binned to $90$ minutes.
    
    The \textrm{MAXI} instrument has two kinds of X-ray cameras, the Gas Slit Camera (GSC) \citep{maxi_gsc_main_paper} and the Solid-state Slit Camera (SSC) \citep{maxi_ssc_main_paper}. The GSC instrument is a large-area position-sensitive proportional counter with a total detection area of $5350\,\mathrm{cm}^2$, operating in the energy range of \text{2$-$30 $\mathrm{keV}$}. SSC has a detection area of $200\,\mathrm{cm}^2$ and operates in an energy range of 0.5$-$12 $\mathrm{keV}$. For this study, we used only the GSC instrument in the energy range of 2$-$20 $\mathrm{keV}$. SSC was not used because of its low detection area and lower available exposure, which lead to a very low overall count rate, even for a bright source such as Cen X-3.
    
    The GSC light curve and spectrum were obtained from the \textrm{MAXI} on-demand process website\footnote{\url{http://maxi.riken.jp/mxondem/index.html}}. We used GSC data ranging from MJD 55075 to MJD 60034, a total duration of $\sim$13.5 years. The effective exposure time of Cen X-3 is approximately $11.39\,\mathrm{Ms}$, or $\sim$$132$ days. The light curves were obtained with a time resolution of $90$ minutes. Orbital-phase-resolved and intensity-resolved data were similarly downloaded using appropriate good time intervals.

\section{Timing analysis} \label{timing_analysis}

    \begin{figure}
    \centering
    \includegraphics[width=0.9\linewidth]{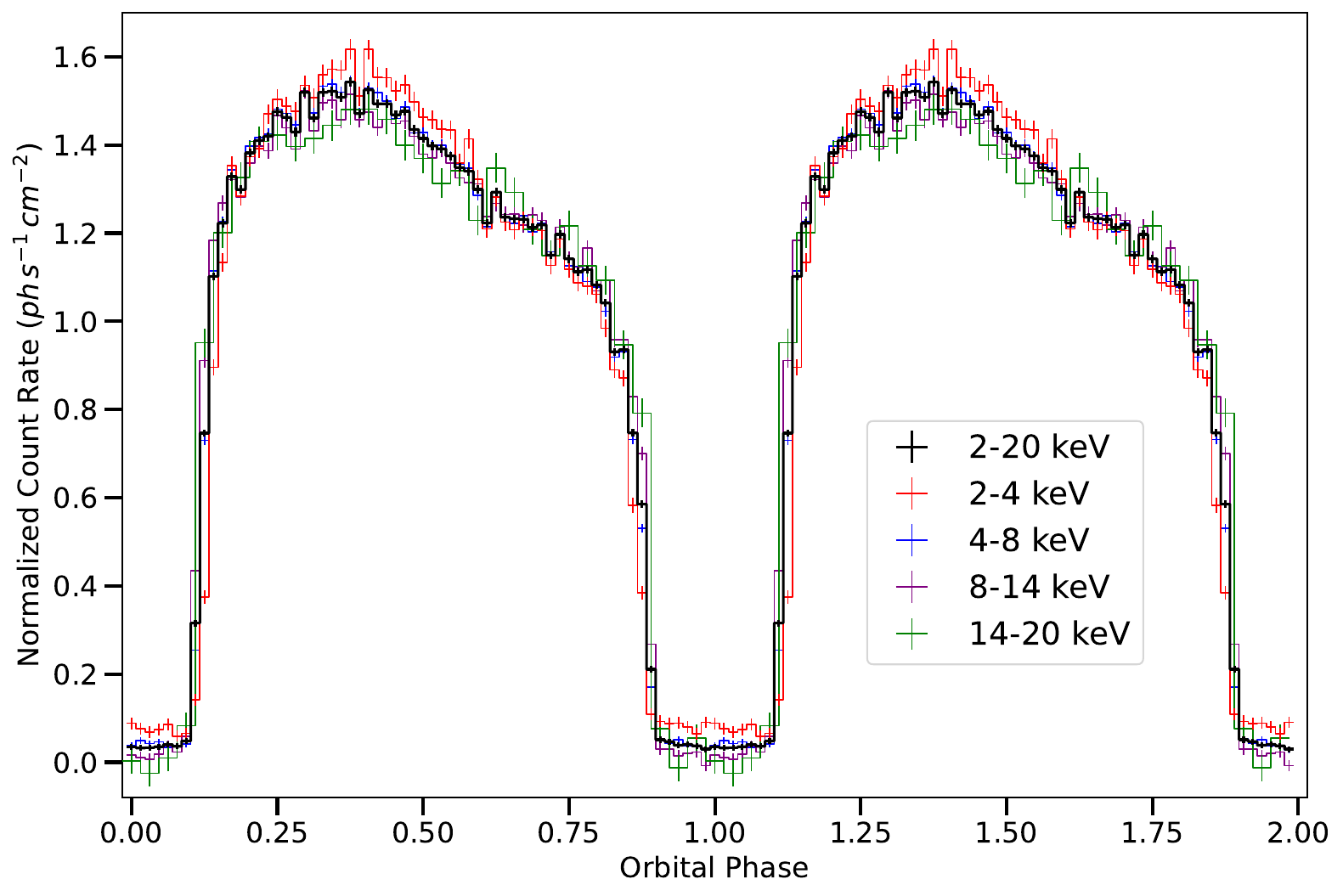}
    \caption{Energy-resolved normalized orbital profile in multiple energy bands of \text{2$-$20 $\mathrm{keV}$}, \text{2$-$4 $\mathrm{keV}$}, \text{4$-$8 $\mathrm{keV}$}, \text{8$-$14 $\mathrm{keV}$}, and \text{14$-$20 $\mathrm{keV}$}. The orbital ephemeris is taken from \cite{moritz_klawin_orbital_ephemeris_2023}.}
    \label{fig:folded_curve}
    \end{figure}
    
    The long-term light curve of Cen X-3 is shown in \text{Figure \ref{fig:long_term_curve}}, with a bin size of $30$ days and presenting count rate modulation over several orbits. Normalized folded light curves of Cen X-3 for different energy bands (\text{2$-$20 $\mathrm{keV}$}, \text{2$-$4 $\mathrm{keV}$}, \text{4$-$8 $\mathrm{keV}$}, \text{8$-$14 $\mathrm{keV}$}, and \text{14$-$20 $\mathrm{keV}$}) were generated (see \text{Figure \ref{fig:folded_curve}}). The normalization of the folded light curve was performed by dividing the count rates in each phase bin by the average source count rate. The orbital ephemeris for Cen X-3 light curve folding is taken from \cite{moritz_klawin_orbital_ephemeris_2023} (see Table \ref{table:orbital_ephemeris}). The orbital period calculated independently using \texttt{efsearch}\footnote{\url{https://heasarc.gsfc.nasa.gov/xanadu/xronos/examples/efsearch.html}} is $2.08697\pm0.00002$ days for the $\sim$13.5 years of data ($\sim$ 2009 to 2023). This is in good agreement with \cite{Harsha_paul_2010_cenx3_orbital_ephemeris} and \cite{moritz_klawin_orbital_ephemeris_2023}. The orbital ephemeris from \cite{moritz_klawin_orbital_ephemeris_2023} was used for folding in a further analysis.
    
    The energy-resolved normalized orbital profiles exhibit visually similar phase variations for all the energy bands considered. But, looking more closely at the orbital profiles for the \text{2$-$4 $\mathrm{keV}$} (red) and \text{14$-$20 $\mathrm{keV}$} (green) energy bands in \text{Figure \ref{fig:folded_curve}}, we see that the spectrum becomes harder around an orbital phase $\sim$$0.6$ when the softer band (red) goes below the harder band (green).
    
\begin{figure}
    \centering
    \includegraphics[width=0.98\linewidth]{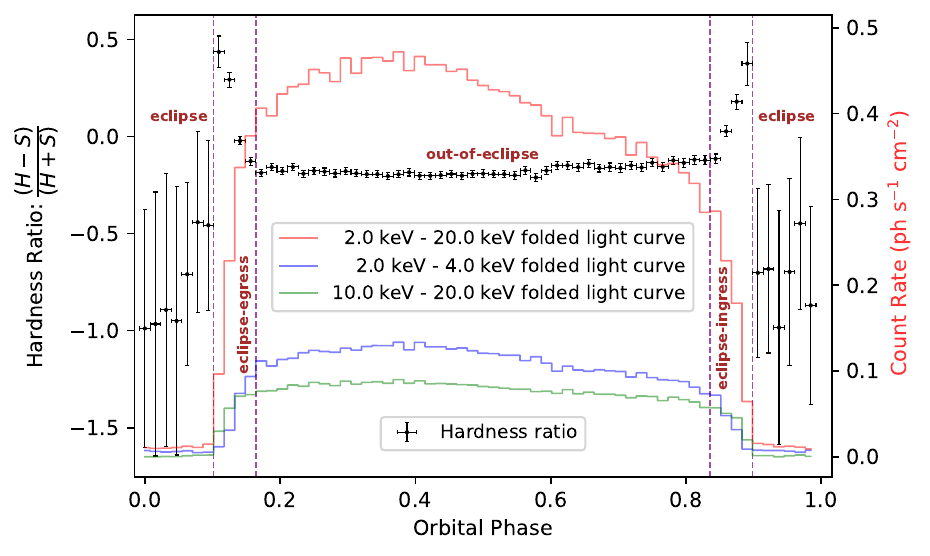}
    \caption{Variation of the HR with the orbital phase as calculated from the energy bands, $S$: 2$-$4 $\mathrm{keV}$ and $H$: 10$-$20 $\mathrm{keV}$. The plot is overlaid with the $S$-band (blue), the $H$-band (green), and the 2$-$20 $\mathrm{keV}$ (red) folded light curve. The vertical dashed purple lines divide the orbit into four regions: out of eclipse (OOE), eclipse ingress, eclipse egress, and eclipse.}
    \label{fig:Hardness_ratio_plots}
    \end{figure}

    To quantitatively study the variation in orbital profile across multiple energy bands, and thus the change in the shape of the source spectrum, we used the $HR$, which is defined as
    \begin{align}
        HR = \dfrac{H-S}{H+S},
    \end{align}
    where $S$ and $H$ are the source count rates in the soft (\text{2$-$4 $\mathrm{keV}$}) and hard (\text{10$-$20 $\mathrm{keV}$}) energy bands, respectively. Here we chose \text{10$-$20 $\mathrm{keV}$} for the harder band (instead of \text{14$-$20 $\mathrm{keV}$} used above) for better statistics, since the count rate from the hard band is closer to the soft band this way. We calculated the variation of the $HR$ across 64 orbital phase bins for an orbit and plotted it in \text{Figure \ref{fig:Hardness_ratio_plots}}. The $HR$ sharply increases before and after the eclipse, indicating an increase in absorption from the stellar wind as the compact object gradually goes behind the companion star. So we defined different regions in orbit based on different values of $HR$, which may originate from a variable absorption column density.
    
    The $HR$ thus serves as a proxy to differentiate between the multiple orbital regions, namely the eclipse ingress when the $HR$ increases sharply, eclipse egress when the $HR$ decreases sharply, out of eclipse (OOE) when the $HR$ stays almost constant for Cen X-3, and eclipse. The eclipse-egress, OOE, and eclipse-ingress regions have orbital phases ranging from $\phi_{\mathrm{orb}}\sim$ 0.102$-$0.164, 0.164$-$0.836, and 0.836$-$0.898, respectively, based on the variation of the $HR$ with the orbital phase as seen in Figure \ref{fig:Hardness_ratio_plots}.
    
\section{Spectral analysis} \label{spectral_analysis}

    All the spectra from \textrm{MAXI}/GSC were analysed using {\texttt{XSPEC} v12.13.0c} \citep{xspec_arnaud_1996}. The elemental abundances used are from \cite{Wilms_2000}, and the photoionization absorption cross-sections are taken from \cite{vern_1996}.
    
    \subsection{Intensity-averaged and orbital-phase-averaged spectrum}
    \label{sec:phase_average_spec}
    
    \begin{figure}
    \centering
    \includegraphics[width=1.\linewidth]{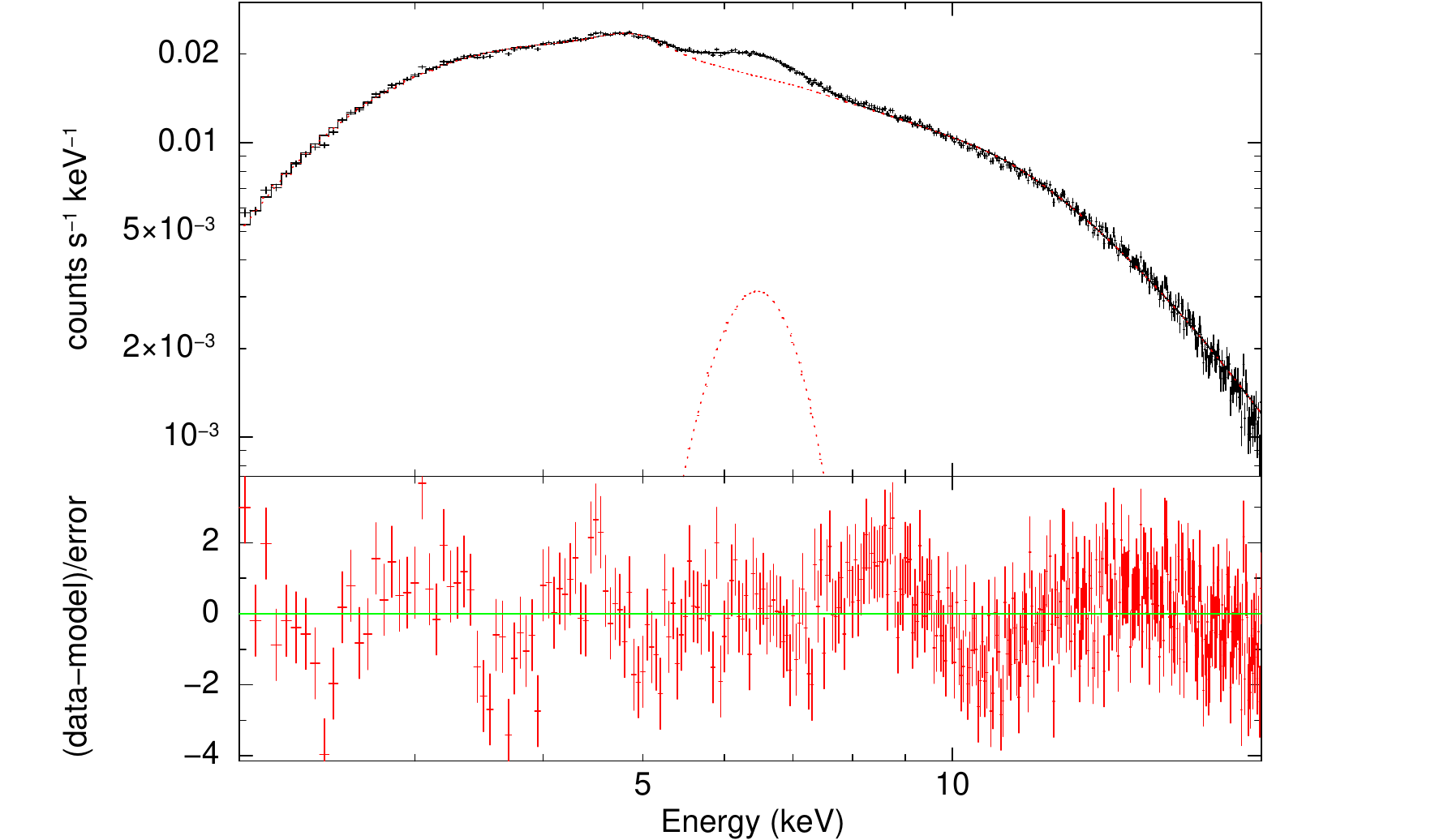}
    \caption{Long-term intensity-averaged and orbital-phase-averaged 2$-$20 $\mathrm{keV}$ spectrum of Cen X-3 for the $\sim$13.5 years of \textrm{MAXI}/GSC data. Top panel: 2$-$20 $\mathrm{keV}$ \textrm{MAXI}/GSC intensity-averaged and orbital-phase-averaged spectrum and the best-fit spectral model of Cen X-3. Bottom panel: Residual of the best-fit spectral model.}
    \label{fig:model_spectrum} 
    \end{figure}
    
    The long-term orbital-phase-averaged 2$-$20 $\mathrm{keV}$ \textrm{MAXI}/GSC spectrum of Cen X-3 is fitted with a partially absorbed power law with a high-energy cutoff continuum model and a Gaussian feature for the iron $\mathrm{K}\alpha$ fluorescent emission line. We used the spectral model  \texttt{TBabs*pcfabs*( powerlaw*highecut + gaussian )}. Cen X-3 has a soft excess, similar to many X-ray pulsars, which is usually fitted with a blackbody component. But the phenomenological spectral model used in this paper does not require an additional blackbody component; the reason for this is elaborated on in the discussion below.
    
    The absorption in X-ray has two components, one due to the circumstellar matter around the binary system and the other due to the Galactic interstellar medium (ISM). The local absorption due to circumstellar matter from stellar wind was modelled with \texttt{pcfabs}, a partial covering absorption model. The absorption due to the Galactic ISM was modelled with \texttt{TBabs}, and the value of \texttt{Tbabs:} nH, the equivalent hydrogen column density, was fixed to the Galactic line-of-sight value of $1.11\times10^{22}\,\mathrm{atoms\,cm}^{-2}$ \citep{nh_value_nasa_HI4PI_Collaboration}. The best-fit values of the spectral parameters for the model used are given in \text{Table \ref{table:intensity_res_spectral_fit}} for the intensity-averaged and orbital-phase-averaged spectrum. The best-fit spectrum is plotted in Figure \ref{fig:model_spectrum}, along with the residuals to the best-fit model. The 2$-$20 $\mathrm{keV}$ flux for the phase-averaged spectrum is $2.89\pm0.01\times10^{-9}\mathrm{\,ergs\,s^{-1}\,cm^{-2}}$.
    
    \subsection{Intensity-averaged and orbital-phase-resolved spectroscopy}
    \label{subsec:orbital_phase_resolved_spectrscopy}
    
    \begin{figure}
    \centering
    \includegraphics[width=1.0\linewidth]{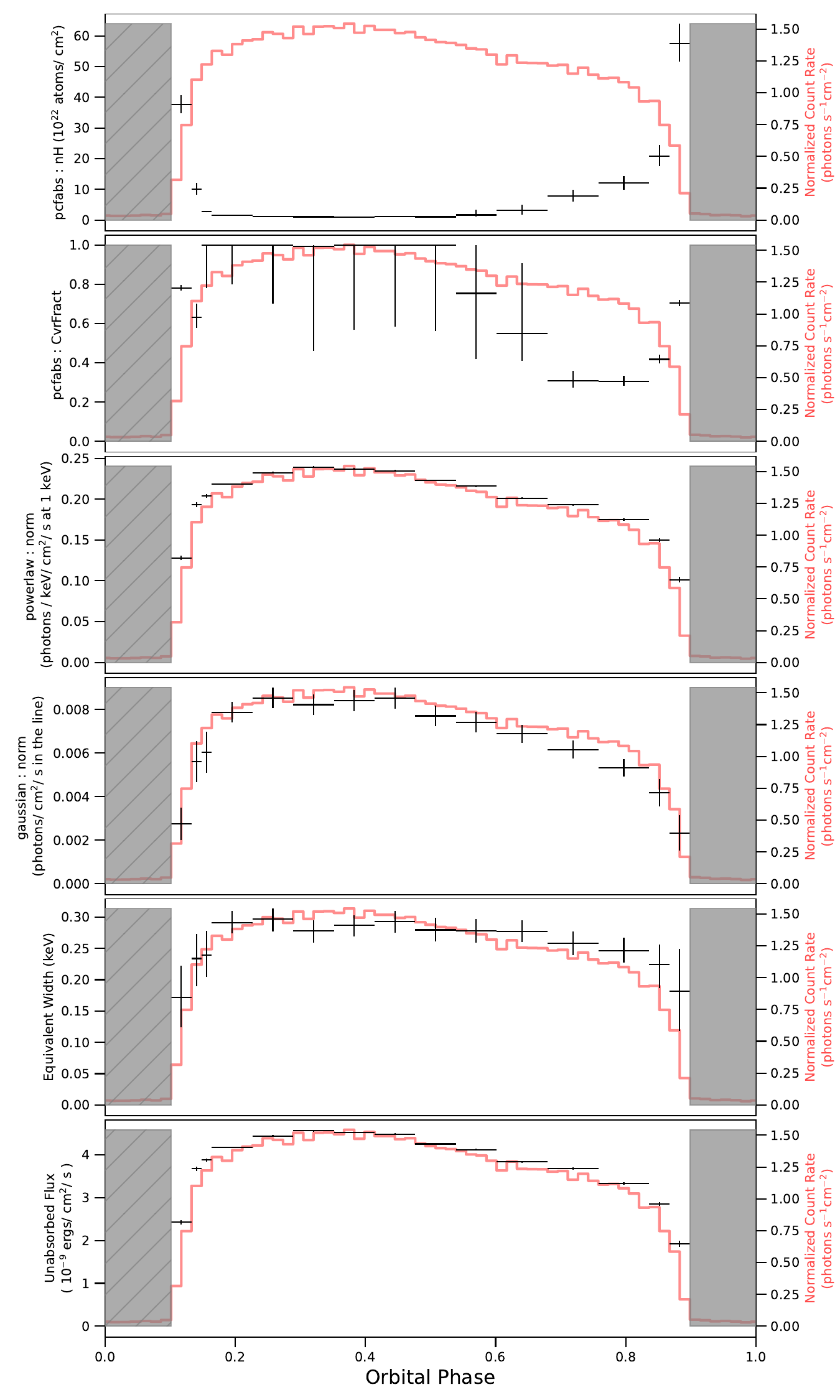}
    \caption{Variations of free spectral parameters plotted from the intensity-averaged orbital-phase-resolved spectral analysis of the \text{2$-$20 $\mathrm{keV}$} \textrm{MAXI}/GSC data. The 2$-$20 $\mathrm{keV}$ unabsorbed flux for the spectral model used is calculated and plotted at the bottom. The errors are quoted at a 1$\sigma$ confidence level. The shaded region indicates the eclipse phase of Cen X-3.}
    \label{fig:phase_resolved_spectral_parameters} 
    \end{figure}
    
    The binary orbit of Cen X-3 is divided into four regions, namely eclipse-egress, OOE, eclipse-ingress, and eclipse, based on the $HR$ variation. The orbital phase gives the relative positions of the NS and the companion star with respect to our line of sight.
    
    Further, the OOE region is divided into ten \text{phase bins}, while the eclipse-ingress and eclipse-egress regions are divided into two and three \text{phase bins}, respectively. The counts of the new phase bins for each orbital phase range were kept as uniform as possible to maintain similar statistics. 
    
    Orbital-phase-resolved spectroscopy was performed with the same spectral model used for the orbital-phase-averaged spectrum on the spectra extracted for the different orbital phase bins. For the spectral fitting in the orbital-phase-resolved spectra, some parameters were frozen to the best-fit values of the spectral model in Table \ref{table:intensity_res_spectral_fit}. The spectral parameters that were frozen are  \texttt{TBabs,} nH; \texttt{powerlaw,} $\Gamma$, \texttt{highecut,} $E_{\mathrm{cutoff}}$; \texttt{highecut,} $E_{\text{fold}}$, \texttt{gaussian,} $E_\mathrm{Fe}$; and \texttt{gaussian,} $\sigma_\mathrm{Fe}$.
    
    Cen X-3 has a highly circular orbit with eccentricity ${<1.6\times10^{-3}}$ \citep{bildsten_cenx3_small_eccentricity, Harsha_paul_2010_cenx3_orbital_ephemeris}; therefore, the emission from the NS is likely to be independent of the orbital phase, as matter density in the circumbinary region used for accretion does not change due to a constant orbital separation between the NS and the optical companion. The accretion rate may still change within or across orbit(s), depending on the behaviour of the circumbinary medium or local conditions near the NS, but it will not have any clear orbital phase dependence in the long term. So only photoelectric absorption of the emission from the NS may change and show an orbital phase dependence. Thus, the continuum parameter $\Gamma$ is frozen, as seen above, to the phase-averaged spectral fit value for orbital-phase-resolved spectroscopy. After the spectral fitting of each phase bin spectrum, we can see the evolution of the free parameters in the binary orbit in ${\text{Figure } \ref{fig:phase_resolved_spectral_parameters}}$.
    
    \begin{figure}
    \centering
    \includegraphics[width=1.0\linewidth]{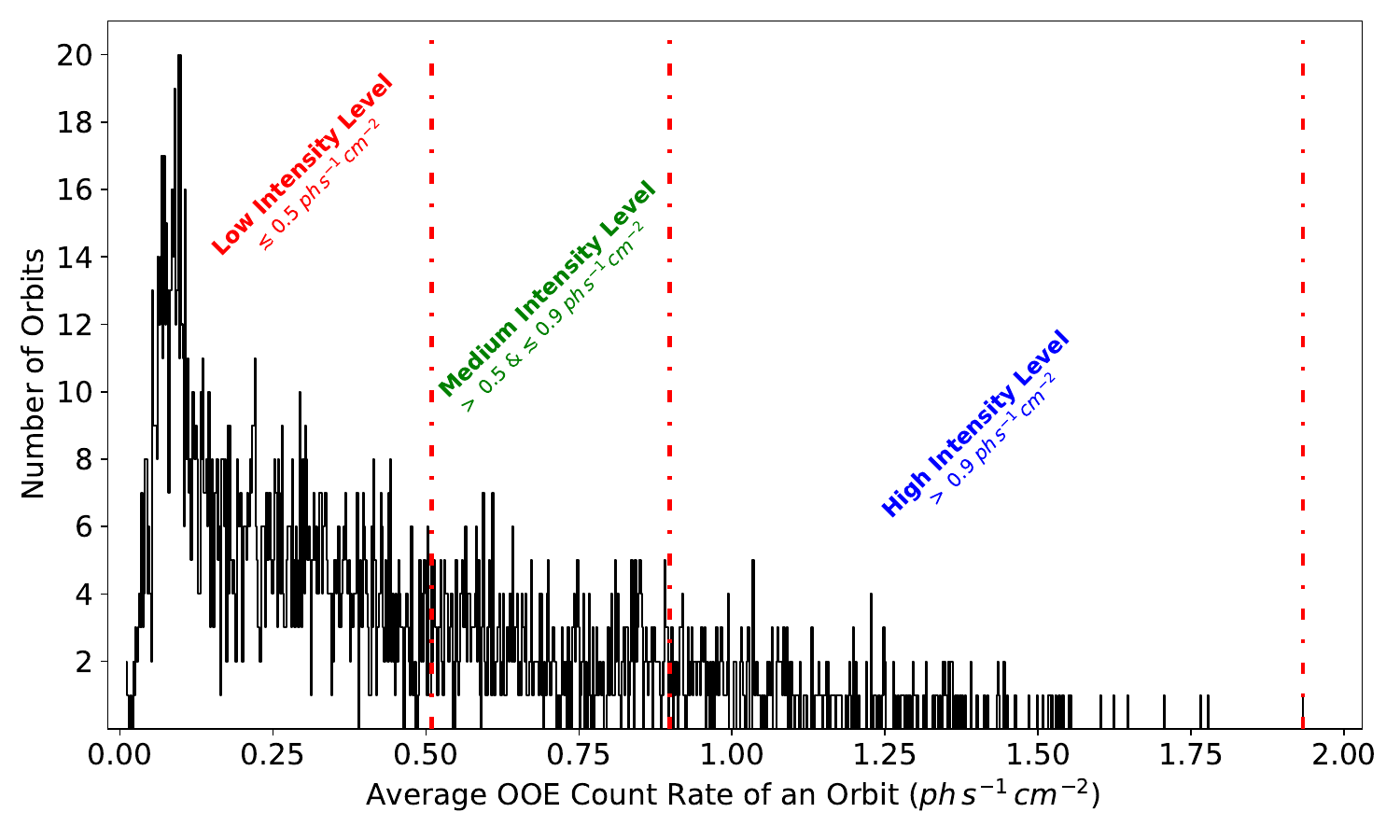}
    \caption{Histogram of total number of orbits vs average out-of-eclipse count rate per orbit, where we divide all the orbits into three intensity levels.}
    \label{fig:histogram_intensity_level} 
    \end{figure}

    \subsection{Intensity-resolved analysis}
    
    \begin{figure}
    \centering
    \includegraphics[width=0.9\linewidth]{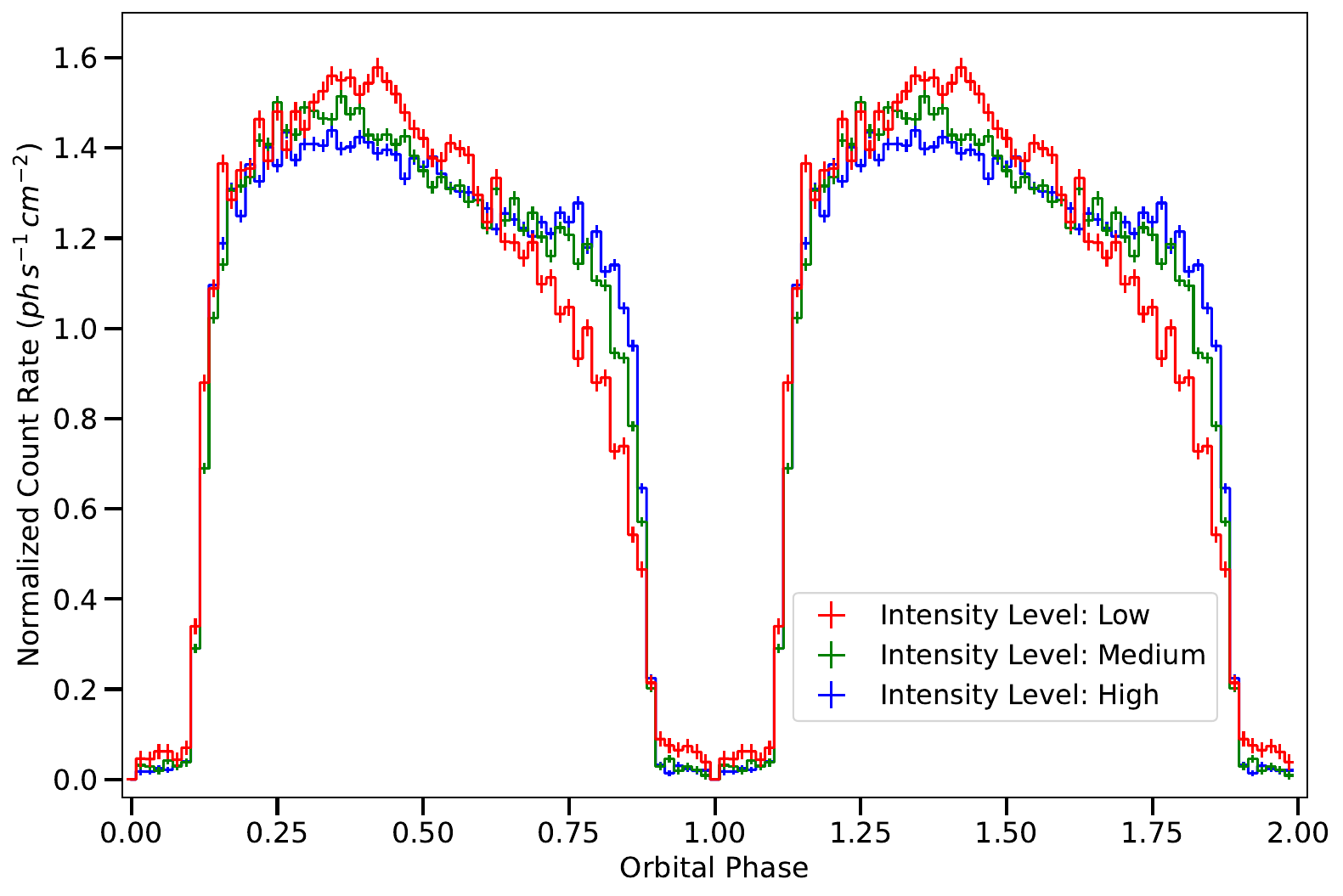}
    \caption{Normalized orbital profile for all three intensity levels.}
    \label{fig:intensity_resolved_study_lc}
    \end{figure}
    
    The orbit-averaged out-of-eclipse count rate from Cen X-3 shows significant orbit-to-orbit variation. From the normalized folded light curve in $\text{Figure}\,\ref{fig:folded_curve}$, we obtained the orbital phase range for the OOE region, which is ${\phi_\text{orb}\sim0.164-0.836}$. Good time intervals for the OOE region were generated according to the orbital ephemeris given in \text{Table \ref{table:orbital_ephemeris}} and used to generate a 2$-$20 $\mathrm{keV}$ \textrm{MAXI}/GSC light curve of Cen X-3 from MJD 55075 to MJD 60034. Then the average count rate for each orbit during the OOE phase was calculated. The variation in this orbit-averaged OOE count rate shows no obvious long-term trend. A histogram for the total number of orbits against the average OOE count rate for each orbit is seen in Figure \ref{fig:histogram_intensity_level}. For this histogram, we redistributed the $\sim$13.5 years of \textrm{MAXI}/GSC data for Cen X-3 into three intensity levels, as shown with the vertical dashed lines in \text{Figure \ref{fig:histogram_intensity_level}}, where the low, medium, and high intensity levels have 1477, 467, and 279 orbits, respectively. This was done to keep the total photon counts similar across all intensity levels in order to maintain similar statistics.

    \subsection{Intensity-resolved and orbital-phase-averaged spectral analysis}
    
    The normalized orbital profile for each intensity level is plotted in $\text{Figure \ref{fig:intensity_resolved_study_lc}}$. The normalized orbital profile for the three intensity levels is significantly different. For the lowest intensity level, the profile is more asymmetric with respect to orbital phase 0.5 compared to the medium and high intensity levels. There is a steep decrease in source count rate beyond the orbital phase $\phi_\text{orb}\sim0.5$ for the lowest intensity level. The orbital-phase-averaged spectrum for each intensity level was fitted with the same spectral model used in Section \ref{sec:phase_average_spec} for the overall spectra. The best-fit parameters for the spectral model for all intensity levels are shown in Table \ref{table:intensity_res_spectral_fit}. The local absorption is smaller for higher intensity levels.
    
    \begin{figure}
    
    \includegraphics[width=0.9\linewidth]{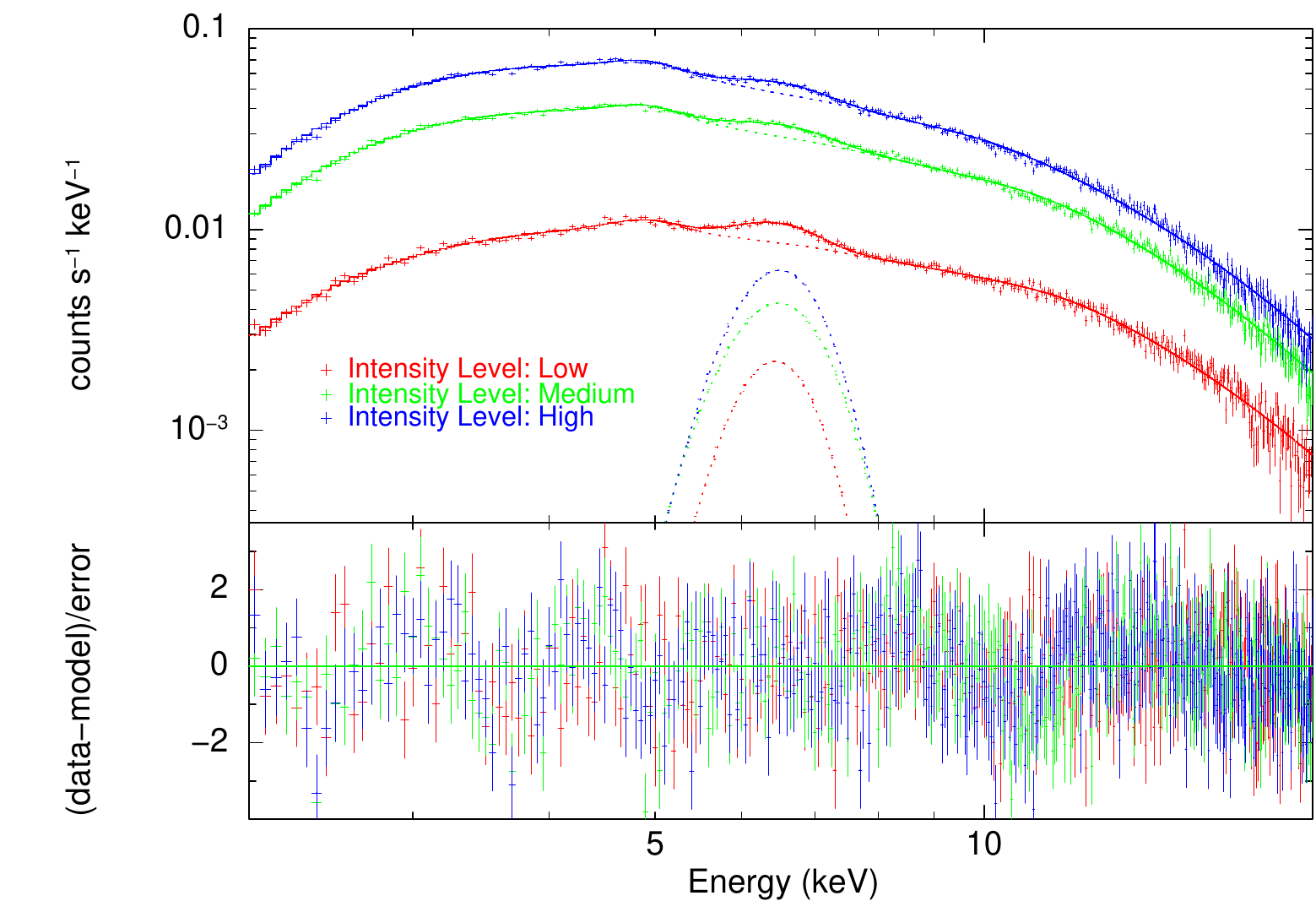}
    \caption{ Intensity-resolved and orbital-phase-averaged 2$-$20 $\mathrm{keV}$ spectrum of Cen X-3 for the $\sim$13.5 years of \textrm{MAXI}/GSC data. Top panel: 2$-$20 $\mathrm{keV}$ \textrm{MAXI}/GSC orbital-phase-averaged spectrum for all three intensity levels of Cen X-3 with their best-fit spectral model. Bottom panel: Residual of the best-fit spectral model for all intensity levels.}
    \label{fig:intensity_resolved_study_spec}
    \end{figure}
    
    \subsection{Intensity-resolved and orbital-phase-resolved spectroscopy}
    
    \begin{figure}
    \centering
    \includegraphics[width=1.0\linewidth]{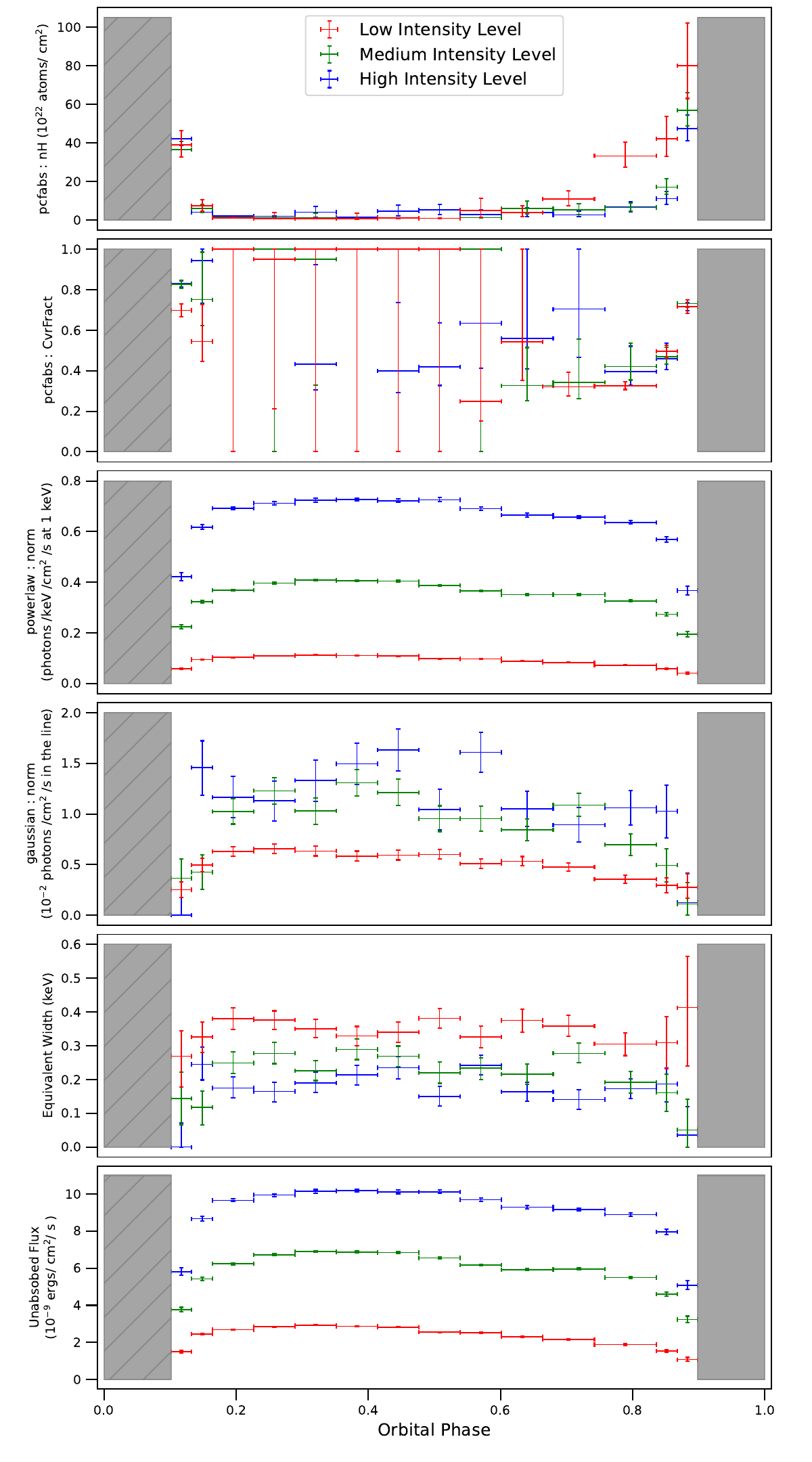}
    \caption{Intensity-resolved and orbital-phase-resolved spectroscopy for all three intensity levels plotted in red, green, and blue, respectively. The 2$-$20 $\mathrm{keV}$ unabsorbed flux for the spectral model used is calculated and plotted at the bottom. The errors are quoted at a 1$\sigma$ confidence level. The shaded region indicates the eclipse phase of Cen X-3.}
    \label{fig:int_res_orb_res_spec_para} 
    \end{figure}
    
    \begin{table*}
    \caption{Best-fit results of the spectral model used for the \text{2$-$20 $\mathrm{keV}$} intensity-averaged and intensity-resolved orbital-phase-averaged spectrum}
    \label{table:intensity_res_spectral_fit}
    \renewcommand{\arraystretch}{1.5}
    \small
    \centering
    \begin{threeparttable}
    \setlength\tabcolsep{8pt}
    \begin{tabular}{llllll}
        \hline
        Component           & Parameter                         &                                           & Best-Fit Values$^{\text{a}}$                  &                                               &                                                   \\
                            &                                   & Intensity Average                         & Low Intensity Level                           & Middle Intensity Level                        & High Intensity Level                              \\ \hline
        \texttt{TBabs}      & nH (10$^{22}$ cm$^{-2}$)          & 1.11                                      & 1.11                                          & 1.11                                          & 1.11                                              \\
                            &                                   & (frozen)                                  & (frozen)                                      & (frozen)                                      & (frozen)                                          \\ \hline
        \texttt{pcfabs}     & nH (10$^{22}$ cm$^{-2}$)          &  6.0 $\pm\,$1.1                           & 19.6 $^{+3.2}_{-3.6}$                         & 3.8 $\pm\,$1.5                                & 3.2 $\pm\,$1.1                                    \\
                            & CvrFract                          & 0.35$^{+0.03}_{-0.02}$                    & 0.27 $\pm\,$0.04                              & 0.44 $^{+0.15}_{-0.06}$                       & 0.6 $^{+0.2}_{-0.1}$                              \\ \hline
        \texttt{powerlaw}   & $\Gamma$                          & 1.12$^{+0.03}_{-0.01}$                    & 0.98 $\pm\,$0.04                              & 1.18 $\pm\,$0.02                              & 1.26 $\pm\,$0.02                                  \\
                            & norm$^{\mathrm{b}}$                 & 0.16 $\pm\,$0.01                          & 0.066 $^{+0.007}_{-0.006}$                    & 0.31 $\pm\,$0.01                              & 0.59 $\pm\,$0.02                                  \\ \hline
        \texttt{highecut}   & $E_\mathrm{cutoff}$ ($\mathrm{keV}$)           & 11.1 $\pm\,$0.1                           & 11.6 $\pm\,$0.2                               & 11.1 $\pm\,$0.2                               & 9.9 $\pm\,$0.2                                    \\
                            & $E_\mathrm{fold}$ ($\mathrm{keV}$)             & 10.7 $\pm\,$0.3                           & 10.4 $\pm\,$0.6                               & 10.5 $\pm\,$0.5                               & 11.7 $\pm\,$0.4                                   \\ \hline
        \texttt{gaussian}   & $E_\mathrm{Fe}$ ($\mathrm{keV}$)             & 6.43 $\pm\,$0.02                          & 6.41 $\pm\,$0.02                              & 6.46 $\pm\,$0.03                              & 6.47 $\pm\,$0.03                                  \\
                            & $\sigma_\mathrm{Fe}$  ($\mathrm{keV}$)       & 0.38 $\pm\,$0.03                          & 0.30 $\pm\,$0.04                              & 0.45 $\pm\,$0.06                              & 0.39 $\pm\,$0.06                                  \\
                            & Equivalent Width ($\mathrm{keV}$)                        & 0.29 $\pm\,$0.01                          & 0.34 $^{+0.03}_{-0.02}$                       & 0.23 $\pm\,$0.02                              & 0.19 $^{+0.02}_{-0.01}$             \\
                            & norm$_\mathrm{Fe}$$^{\mathrm{c}}$   & 5.5 $\pm\,$0.2 $\times\,$10$^{-3}$        & 3.7 $\pm\,$0.2 $\times\,$10$^{-3}$            & 8.1 $\pm\,$0.6 $\times\,$10$^{-3}$            & 11.1 $\pm\,$0.9 $\times\,$10$^{-3}$           \\ \hline
                            & Unabsorbed Flux$^{\mathrm{d}}$      & 3.09 $\pm\,$0.01 $\times\,$$10^{-9}$    & 1.72 $\pm\,$0.03 $\times\,$10$^{-9}$          & 5.24 $\pm\,$0.03 $\times\,$10$^{-9}$   & 8.30 $\pm\,$0.05 $\times\,$10$^{-9}$    \\ \hline
                            & {$\chi^{2}_\mathrm{red}$ (d.o.f.)}   & $\sim$1.55\,(350)                         & $\sim$1.19\,(347)                             & $\sim$1.26\,(350)                             & $\sim$1.39\,(350)                                 \\ \hline
    \end{tabular}
    \begin{tablenotes}
        \small
        \item[$\text{a}$] {\scriptsize The error bar is quoted at a 68.27$\%$ confidence limit.}
        \item[$\text{b}$] {\scriptsize in units of photons $\mathrm{keV}^{-1}$ s$^{-1}$ cm$^{-2}$ at 1 $\mathrm{keV}$.}
        \item[$\text{c}$] {\scriptsize in units of total photons s$^{-1}$ cm$^{-2}$ in the line.}
        \item[$\text{d}$] {\scriptsize unabsorbed flux found using \texttt{cflux} within the 2$-$20 $\mathrm{keV}$ range for the spectral model used in units of ergs s$^{-1}$ cm$^{-2}$.}
    \end{tablenotes}
    \end{threeparttable}
\end{table*}
    
    For each intensity level, the orbital profile was divided into two eclipse-egress, ten OOE, and two eclipse-ingress phase bins. The spectrum for each phase bin was then fitted with the best-fit spectral model for the intensity-resolved and the orbital-phase-averaged spectrum for a given intensity level. Just as in Section \ref{subsec:orbital_phase_resolved_spectrscopy}, we fixed the absorption due to Galactic ISM (\texttt{TBabs:} nH), the photon index ($\Gamma$), the high-energy cutoff ($E_\mathrm{cutoff}$ and $E_\mathrm{fold}$), the Gaussian line centre ($E_\mathrm{Fe}$), and the line width ($\sigma_\mathrm{Fe}$) for the $\mathrm{Fe}$ $\mathrm{K}_\alpha$ fluorescent emission line. The orbital variation in the free spectral parameter for each intensity level is plotted in Figure \ref{fig:int_res_orb_res_spec_para}.
    
    The local photoelectric absorption increases significantly after $\phi_{\mathrm{orb}}\sim0.5$ for the lowest intensity level. Even though there is some excess absorption present at the high and medium intensity levels with respect to a symmetric wind model, there exists a significant deviation in absorption column density for the lowest intensity level, indicating a significant increase in absorbing matter along the line of sight for the source.

\section{Wind modelling} \label{sec:wind_modelling}

    \subsection{Intensity-averaged wind modelling}
    The variation in local absorption (\texttt{pcfabs}: nH) observed in \text{Figure \ref{fig:phase_resolved_spectral_parameters}} is not a symmetric function of $\phi_{\mathrm{orb}}$. The local absorption would have been a symmetric function of $\phi_{\mathrm{orb}}$ for spherically symmetric stellar wind. There is excess absorption after $\phi_{\mathrm{orb}}\sim0.5$, which may originate from an asymmetric structure present in the binary environment of Cen X-3. Such a feature after $\phi_{\mathrm{orb}}=0.5$ is attributed to the presence of matter that trails behind the NS or is disrupted by it \citep{ blondin_wind_sim_1990, wind_mod_ex_footprints_velax1_paper}. Also, the photoelectric absorption is larger at pre-eclipse than at post-eclipse.
    
    \begin{figure}
    \centering
    \includegraphics[width=1.0\linewidth]{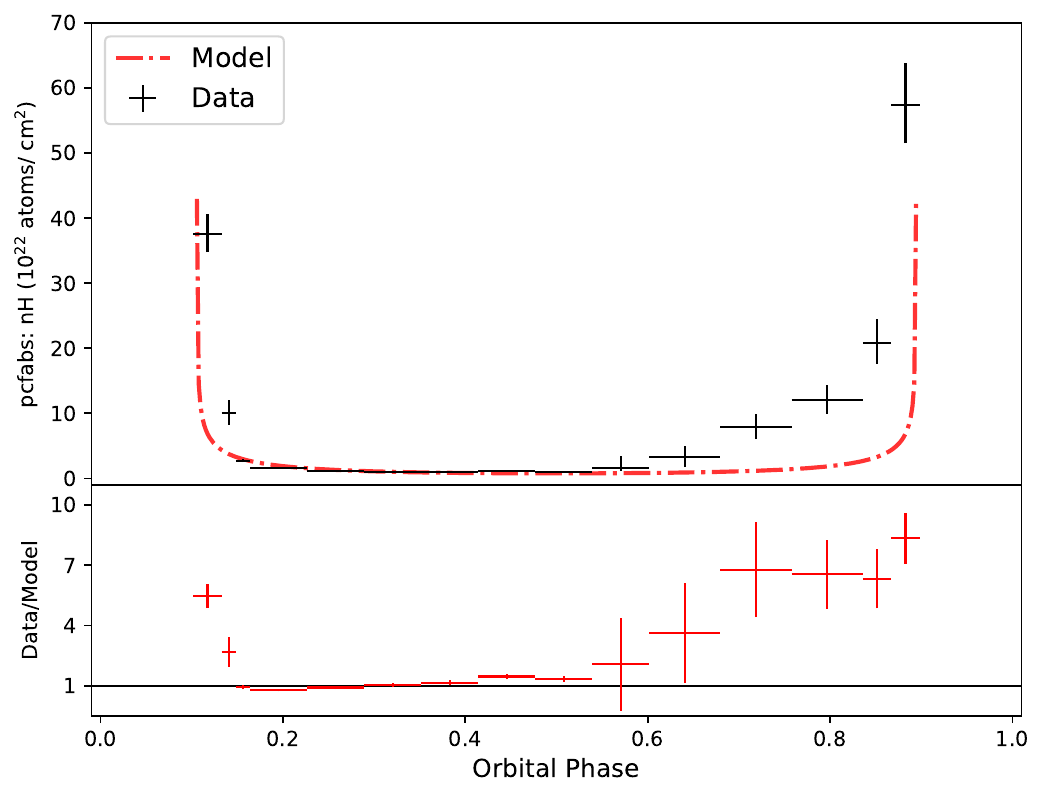}
    \caption{Local absorption evolution with the orbital phase. Top panel: Variation in local absorption across the orbit overlaid with the best-fit spherically symmetric wind model for the intensity-averaged data. Bottom panel: Ratio between the observed absorption and the best-fit model.}
    \label{fig:wind_model_fit_intensity_average} 
    \end{figure}
    
    The variation in photoelectric absorption with orbital phase in HMXBs is usually modelled with a spherically symmetric, radiatively driven wind model \citep{wind_mod_ex_pulse_phase_res_anal_cenx3_2_orb,wind_mod_ex_footprints_velax1_paper,wind_mod_ex_4U_1538_52_with_MAXI}. The wind model used is from \cite{CAK_75_wind_model} (CAK75). The CAK75 velocity profile of the stellar wind is given by
    \begin{align}
        v(r) = v_\infty\left(1 - \dfrac{R_\star}{r}\right)^\beta,
    \end{align}
    where $v_\infty$ is the terminal velocity of the wind, $R_\star$ is the radius of the companion star, and $\beta$ is the velocity gradient parameter. The mass-loss rate due to stellar wind from the companion star is $\Dot{m}$, and the radial mass density profile of the stellar wind is given by
    \begin{align}
        \rho(r) = \dfrac{\Dot{m}}{4{\pi}r^2v(r)} = \left(\dfrac{\Dot{m}}{v_\infty}\right)\dfrac{1}{4{\pi}r^2}\dfrac{1}{(1-{R_\star}/{r})^\beta}.
    \end{align}
    The hydrogen mass density is ${\rho{_{_\text{H}}}(r)=x{_{_\text{H}}}\rho(r)}$, where $x_{_\text{H}}$ is the hydrogen mass fraction for abundance taken from \cite{Wilms_2000} during the spectral analysis above. Finally, the equivalent hydrogen number column density, $N_\text{H}$, was calculated by integrating ${\rho{_{_\text{H}}}(r)}$ from the position of the NS to the observer along the line of sight as
    \begin{align}
        N_{_\text{H}}(\phi') = \kappa \int_{x_{\phi'}}^{\infty} {\rho_{_\text{H}}(r(x',y_{\phi'},z_{\phi'}))} \, dx',
    \end{align}
    where $\kappa$ is the conversion factor from the mass to hydrogen particle number, the x-axis is parallel to the line of sight, and $r(x_{\phi'},y_{\phi'},z_{\phi'})$ is the radial distance of the NS from the centre of the companion star at orbital phase $\phi'$. The radius of the companion star in \text{Cen X-3} is taken as ${R_\star=12.1\pm0.5\,R_\odot}$, the orbital radius as $r_d=19.1_{-0.5}^{+0.6}\,R_\odot$ \citep{van_der_meer_2007}, and the orbital inclination angle as ${i=79\pm3^\circ}$ \citep{cenx3_inclination_paper}.
    
    The wind model parameter $\beta$ was fixed to 0.8 \citep{friend_abbott_alpha_0_8} before we fitted the model and calculated the ratio as $\Dot{m}\,(10^{-6}\,M_{\odot}\,\mathrm{yr}^{-1})/v_\infty \,(1000\,\mathrm{km\,s}^{-1})=0.33\pm0.04$. It is clear in Figure \ref{fig:wind_model_fit_intensity_average} that the spherically symmetric wind model is not adequate for explaining the asymmetry found with respect to $\phi_{\mathrm{orb}}=0.5$. The photoelectric absorption due to local matter starts rising earlier, after $\phi_{\mathrm{orb}}=0.5$, and deviates from a spherically symmetric wind model, as clearly seen from the ratio variation in the bottom panel of Figure \ref{fig:wind_model_fit_intensity_average}. Such a rise in absorption could be due to matter trailing behind the NS, called a wake \citep{blondin_wind_sim_1990, wind_loss_hydrodynamical_sim_2012}.
    
    \begin{figure}
    \centering
    \includegraphics[width=1.0\linewidth]{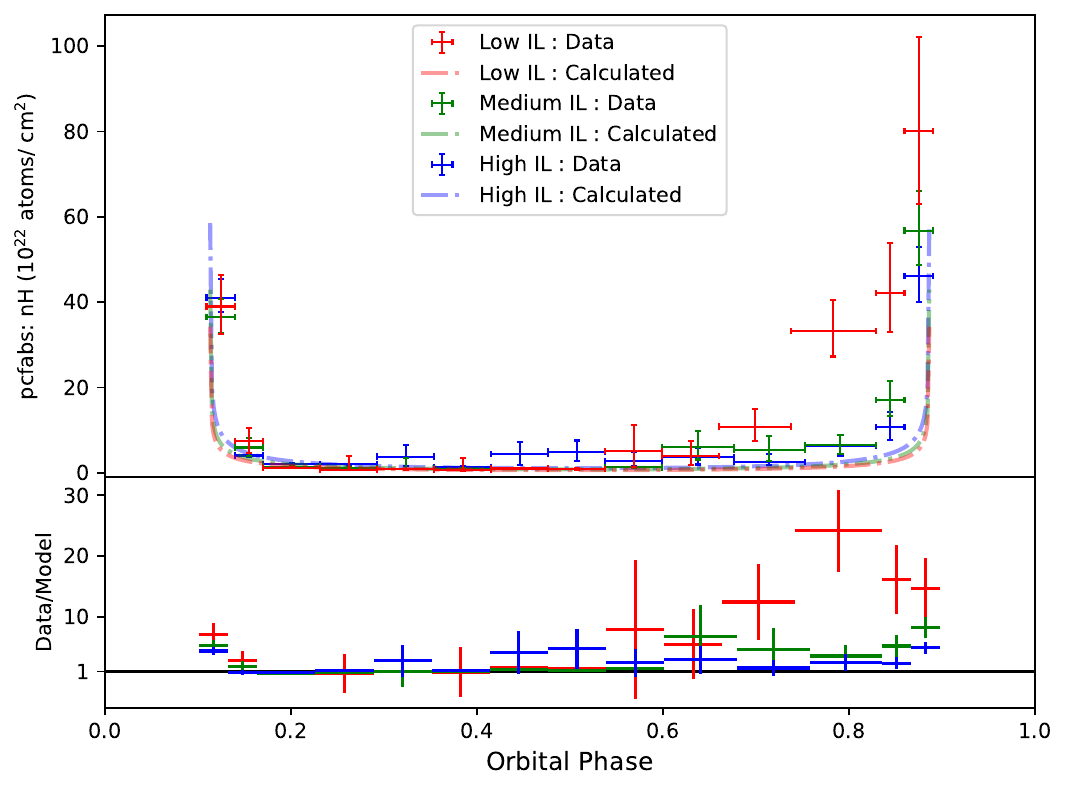}
    \caption{ Intensity resolved local absorption evolution with orbital phase. Top panel: The variation in local absorption with orbital phase for the observed data and the best-fit curve for a spherically symmetric wind model for each intensity level. Bottom panel: The variation in the ratio of data and best-fit model curve for each intensity level.}
    \label{fig:int_res_all_int_level_wind_model} 
    \end{figure}
    \subsection{Wind model for different intensity levels}
    
    For this section, we checked if the orbital phase dependence of photoelectric absorption at each intensity level follows a similar trend as that seen in the intensity average case. We used the CAK75 wind model for all intensity levels. With $\beta=0.8$ \citep{friend_abbott_alpha_0_8} fixed, we get the best-fit values of $\Dot{m}/v_\infty$ for all intensity levels. We determined that the ratio $ \Dot{m}\,(10^{-6}\,M_{\odot}\,\mathrm{yr}^{-1})/v_\infty \,(1000\,\mathrm{km\,s}^{-1})$ is $0.26\pm0.06$, $0.33\pm0.05$, and $0.45\pm0.06$ for the low, medium, and high intensity levels, respectively. The variation in photoelectric absorption for both the data and the best-fit wind model is plotted in Figure \ref{fig:int_res_all_int_level_wind_model} for each intensity level. 
    
    \subsection{Wake toy model for the low intensity level}
    
\begin{figure}
    \centering
    \includegraphics[width=0.5\linewidth]{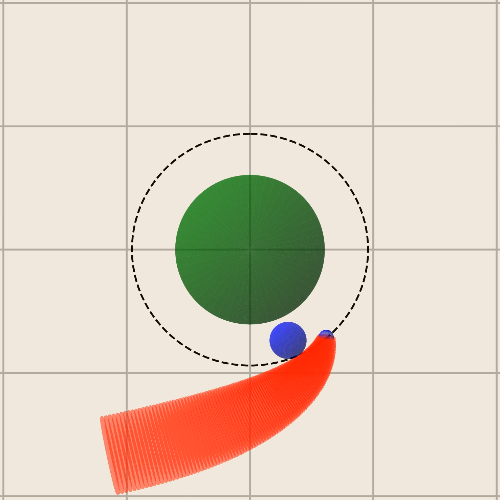}
    \caption{Illustration showing the top view of Cen X-3 with the optical companion in green at the centre along with the dashed line for the orbit of the NS. The structure in red is the accretion wake with a blue tip for the bow shock. The smaller sphere between the dashed line and optical companion is dense matter lagging behind the NS, as a substitute for a photoionization wake or tidal stream.}
    \label{fig:binary_plot_accretion_photoionization_tidal} 
    \end{figure}

\begin{figure}
    \centering
    \includegraphics[width=1.0\linewidth]{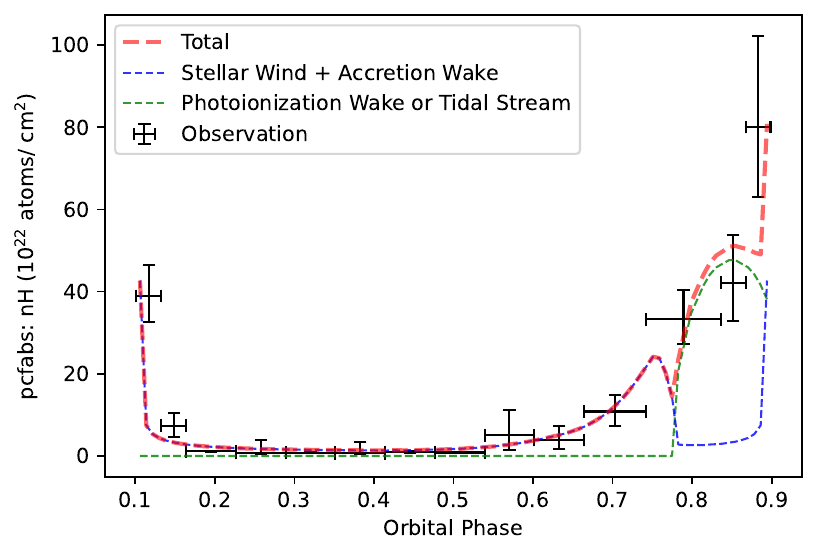}
    \caption{Variation in the observed local absorption with an orbital phase for the low intensity level. Also, absorption due to a stellar wind and other structures, such as an accretion wake, photoionization wake, or tidal stream, calculated for a toy model, is overlaid.}
    \label{fig:accretion_photoionization_tidal} 
    \end{figure}

    The presence of a wake structure trailing the NS at the low intensity level is evident from the increased absorption after $\phi_\mathrm{orb}\sim0.5$. We used a toy model for the accretion wake in Cen X-3 \citep{para_accretion_wake_jackson, wind_mod_ex_pulse_phase_res_anal_cenx3_2_orb}, using a hemisphere for the bow shock, followed by a wake that extends $\sim$$40\,R_\odot$ (see \text{Figure \ref{fig:binary_plot_accretion_photoionization_tidal}}). This toy model has a density of $\sim{3\times10^{-13}\mathrm{\,g\,cm}^{-3}}$ on the boundary and decreases to null towards the inside of the wake and bow shock as a cosine, though accretion wakes should have thin and dense boundaries \citep{fransson_shocks_in_stellar_winds}. Such an assumption works for this demonstration. Along with this, a photoionization wake \citep{fransson_shocks_in_stellar_winds, blondin_wind_sim_1990, photoionization_wake_manousakis_2011} trailing the NS or a tidal stream \citep{Blondin_1991} between the optical star and NS can explain the late-orbital-phase absorption seen in a binary system. A sphere of uniform density $\sim{3\times10^{-12}\mathrm{\,g\,cm}^{-3}}$, an order higher than the accretion wake and lagging behind the NS in \text{Figure \ref{fig:binary_plot_accretion_photoionization_tidal}}, was used in place of a photoionization wake or tidal stream. This explains qualitatively the photoelectric absorption observed during the late orbital phase at the low intensity level of Cen X-3, which can be caused by a photoionization wake or tidal stream that lags behind the NS \citep{fransson_shocks_in_stellar_winds, Blondin_1991}. The absorption due to stellar wind and an accretion wake gets an initial bump between $\phi_\mathrm{orb}\sim$ 0.5$-$0.8, followed by higher absorption before eclipse due to a photoionization wake or tidal stream that results in dense matter lagging behind the NS (see \text{Figure \ref{fig:accretion_photoionization_tidal}}). We note that this is mostly for illustrative purposes, to understand qualitatively the late-orbital-phase absorption at the low intensity level of Cen X-3.

\section{Discussion}
    
    We conducted a long-term study of Cen X-3 and probed the variation in spectra across the orbit to better understand the X-ray absorption characteristics of Cen X-3.
    
    We used a phenomenological spectral model for our spectral analysis, which is different from the $\sim$3 $\mathrm{keV}$ blackbody component used along with the continuum in other reports \citep{cenx3_inclination_paper, cenx3_six_year_paper}. However, the broadband spectrum of Cen X-3 observed with the BeppoSAX \citep{bepposax_burderi_0.1kev_blackbody_2000}, XMM-Newton \citep{blackbody_xmm_broadband}, and Suzaku \citep{blackbody_suzaku_broadband_cenx3_gunjan} observatories, covering a lower energy range than \textrm{MAXI}/GSC, has a blackbody temperature of \text{$\sim$$0.1\,\mathrm{keV}$} and did not require a \text{$\sim$3 $\mathrm{keV}$} blackbody component. A low-temperature blackbody component, such as those at $\sim$$0.1\,\mathrm{keV}$ seen in many other X-ray pulsars, would not have significant emission in the 2$-$20 $\mathrm{keV}$ energy band of \textrm{MAXI}/GSC used in this work.
    
    Instead, we used a partially absorbed power-law continuum with a high-energy cutoff and a Gaussian feature for the strong iron K-alpha fluorescent emission line. For this spectral model, we clearly observed a variation in hydrogen column density, and thus photoelectric absorption, across the binary orbit. Such a variation in absorption helps us understand the characteristics of the absorbing matter in the circumbinary environment of the Cen X-3 binary system.
    
    The variation of photoelectric absorption and the evidence of a wake or matter trailing the NS is clear for the intensity-averaged and orbital-phase-resolved spectra from Figure \ref{fig:wind_model_fit_intensity_average}. We used a simple spherically symmetric wind model to fit the variation in observed photoelectric absorption. Though such a model is not suited for modelling the asymmetry in absorption, we can see how far the data deviate from the model. A clear deviation of the observed absorption away from the symmetric wind model is seen in the evolution of the ratio in the bottom panel of Figure \ref{fig:wind_model_fit_intensity_average}. However, we cannot tell if such a feature is maintained for all the orbits in the long-term data of Cen X-3. So, all the orbits of Cen X-3 were divided into three intensity levels to conduct a more thorough analysis of the asymmetry present in the photoelectric absorption around the binary orbit.
    
    An orbital-phase-resolved spectral analysis was carried out for each intensity level. Clearly, an early rise after $\phi_{\mathrm{orb}}\sim0.5$ is observed, along with very high photoelectric absorption before eclipse for the lowest intensity levels compared to higher intensity levels (see $\text{Figure \ref{fig:int_res_orb_res_spec_para}}$). When we used a symmetric wind model as in the intensity average case, the deviation between the wind model and data for each intensity level was clearly not similar (see $\text{Figure \ref{fig:int_res_all_int_level_wind_model}}$). The photoelectric absorption at the highest intensity level is the most symmetric with respect to $\phi_{\mathrm{orb}}\sim0.5$, as seen in the ratio in Figure \ref{fig:int_res_all_int_level_wind_model}. Meanwhile, the asymmetry in absorption with respect to ${\phi_{\mathrm{orb}}\sim0.5}$ seems to be more prevalent at the lowest intensity level. This suggests that the presence of a wake structure that trails behind the NS is expected for the lowest intensity level of Cen X-3.
    
    A greater value of $L_x/\dot{M}_w$ is expected to lead to a situation where the wind is highly ionized, thus forming a photoionization wake that increases absorption after ${\phi_{\mathrm{orb}}\sim0.5}$ until the eclipse \citep{blondin_wind_sim_1990, Blondin_1991}, where $L_x$ is the X-ray luminosity and $\dot{M}_w$ is the stellar wind mass-loss rate. Assuming that $\dot{M}_w$ is similar across all intensity levels and the late-orbital-phase absorption at the low intensity level of Cen X-3 is due to a photoionization wake, that would mean that a photoionization wake forms for the low intensity level with the lowest $L_x$ or $L_x/\dot{M}_w$ value. This is in contrast with \cite{blondin_wind_sim_1990}, as the high intensity level in Cen X-3 shows the most symmetric absorption with respect to mid-phase, suggesting that a photoionization wake is not as prominent at the high intensity level of Cen X-3.
    
    A structure such as a tidal stream was described by \cite{Blondin_1991} to be steady in time and produce absorption after ${\phi_{\mathrm{orb}}\sim0.5}$ until the eclipse. They also implied that a strong absorption for orbital phase ${\phi_{\mathrm{orb}}\sim0.6-0.7}$ is expected for a significant tidal stream in their model. But we lack such strong absorption for that orbital phase for all intensity levels for Cen X-3 (see $\text{Figure \ref{fig:int_res_all_int_level_wind_model}}$). A Suzaku observation of Cen X-3 at a low intensity level \citep{suzaku_cenx3_paul_naik} has shown extended dips in the X-ray light curve before the eclipse, which indicates X-ray absorption due to dense matter along our line of sight. Dedicated observations of Cen X-3 at different intensity levels with pointed instruments can shed further light on the wake formation in this intriguing binary system.
    
\begin{acknowledgements}
    This research has made use of \textrm{MAXI} data provided by RIKEN, JAXA and the \textrm{MAXI} team. We thank the referee for the valuable comments which helped improve the quality of this paper.
\end{acknowledgements}



\bibliographystyle{aa}
\bibliography{aa51922-24corr} 

\end{document}